\documentclass[aps,pre,twocolumn,a4paper,superscriptaddress,floatfix,10pt,longbibliography]{revtex4-2}

\usepackage[T1]{fontenc}
\usepackage{graphicx}
\usepackage{dcolumn}
\usepackage{bm}
\usepackage{color}
\usepackage{amsmath}
\usepackage{amssymb}
\usepackage{hyperref}
\usepackage{subfigure}
\usepackage{float}
\usepackage{bigints}
\usepackage{changes}
\usepackage{braket}
\usepackage{xcolor}

\newcommand{\X}{\mathit{\boldsymbol{x}}}

\newcommand{\beq}{\begin{equation}}
\newcommand{\eeq}{\end{equation}}
\newcommand{\ba}{\begin{array}}
\newcommand{\ea}{\end{array}}
\newcommand{\bea}{\begin{eqnarray}}
\newcommand{\eea}{\end{eqnarray}}

\newcommand\underrel[2]{\mathrel{\mathop{#2}\limits_{#1}}}

\begin{document}

\title{Zero-temperature Monte Carlo simulations of two-dimensional quantum spin glasses guided by neural network states}

\author{L. Brodoloni}
\affiliation{School of Science and Technology, Physics Division, Universit\`a di Camerino, 62032 Camerino, Italy}
\affiliation{INFN-Sezione di Perugia, 06123 Perugia, Italy}

\author{S. Pilati}
\affiliation{School of Science and Technology, Physics Division, Universit\`a di Camerino, 62032 Camerino, Italy}
\affiliation{INFN-Sezione di Perugia, 06123 Perugia, Italy}

\begin{abstract}
A continuous-time projection quantum Monte Carlo algorithm is employed to simulate the ground state of a short-range quantum spin-glass model, namely, the two-dimensional Edwards-Anderson Hamiltonian with transverse field, featuring Gaussian nearest-neighbor couplings.
We numerically demonstrate that guiding wave functions based on self-learned neural networks suppress the population control bias below modest statistical uncertainties, at least up to a hundred spins.
By projecting a two-fold replicated Hamiltonian, the spin overlap is determined. A finite-size scaling analysis is performed to estimate the critical transverse field where the spin-glass transition occurs, as well as the critical exponents of the correlation length and the spin-glass susceptibility. For the latter two, good agreement is found with recent estimates from the literature for different random couplings. 
We also address the spin-overlap distribution within the spin-glass phase, finding that, for the workable system sizes, it displays a non-trivial double-peak shape with large weight at zero overlap.
\end{abstract}

\maketitle

\section{Introduction}
Despite decades of research, Ising spin glasses still pose intriguing unanswered questions, such as the fate of replica symmetry breaking in finite dimensional lattices, i.e., beyond the mean-field Sherrington-Kirkpatrick (SK) model featuring all-to-all couplings, or the existence of the Almeida-Thouless line in a longitudinal field~\cite{RevModPhys.58.801,charbonneau2023spin}.
In recent years, increasing attention has been devoted to quantum random Ising models, since understanding their zero-temperature properties is instrumental to ascertain the potential efficiency of quantum annealers~\cite{tanaka2017quantum,Hauke_2020}. 
However, accurately simulating the ground state of quantum spin models with frustrated interactions still represents a remarkable computational challenge~\cite{PhysRevB.109.L140408}.

Most computational studies on quantum spin glasses addressed finite-temperature properties, often using path integral Monte Carlo (PIMC) algorithms. The finite-temperature spin-glass transition of the quantum SK model at a weak transverse field was studied in Refs.~\cite{PikYinLai1990,PhysRevE.92.042107}. Some properties of the quantum phase transition were extracted approaching the zero-temperature limit, considering  two- and three-dimensional lattices~\cite{PhysRevLett.72.4141,PhysRevLett.72.4137} (with finite imaginary-time step models), the (quantum) SK model~\cite{Alvarez}, and the Bethe lattice~\cite{Mossi2017}.
The quantum SK model with longitudinal field has been investigated also via a mapping to a one-dimensional model in the thermodynamic limit~\cite{PhysRevE.96.032112} and using a continuous-time PIMC algorithm in presence of full replica symmetry breaking~\cite{kiss2023exact}.

However, PIMC algorithms become computationally expensive in the zero-temperature limit. In fact, only very recently extreme-scale PIMC simulations managed to accurately identify the system-size scaling of the first energy gap, which determines the computational complexity of quantum-annealing optimization~\cite{bernaschi2023quantum}. 
Analyzing the zero-temperature quantum critical properties is possible~\cite{king2023quantum,bernaschi2023quantum}, but it requires effectively reducing a two-dimensional scaling function to a one-dimensional form.
For few-spin models, exact diagonalization methods represent a suitable alternative  to study ground state properties~\cite{DavidLancaster_1997,PhysRevE.92.042107}. Variational ansatzes for the ground-state wave function have also been adopted, specifically, using generalized coherent states for the quantum SK model~\cite{PhysRevLett.129.220401}.

Projection quantum Monte Carlo (PQMC) algorithms represent a suitable choice to simulate the ground states of general quantum many-body systems~\cite{becca2017quantum}. They are potentially exact, at least in the absence of negative sign problems. However, the necessity to control the population of random walkers sometimes leads to systematic biases~\cite{PhysRevA.97.032307,PhysRevB.98.085102,PhysRevE.86.056712,PhysRevB.105.235144,PhysRevB.103.155135}. The population-control bias can be avoided with a sufficiently accurate ansatz for the ground-state wave function, which is used to guide the simulation via importance sampling. In recent years, wave functions based on neural networks have emerged as flexible and accurate ansatzes for variegate quantum many-body systems~\cite{doi:10.1126/science.aag2302,CARLEO2019100311,10.21468/SciPostPhysCodeb.2}. In fact, they have been adopted also as guiding functions in PQMC simulations~\cite{PhysRevB.98.235145,ren2023towards,wilson2021simulations}, eventually implementing self-learning protocols resorting to unsupervised learning from random-walker populations~\cite{PhysRevE.100.043301,PhysRevE.101.063308}. This allows avoiding (sometimes problematic) variational parameter optimizations. However, this self-learning PQMC method has been applied only to one-dimensional nearest-neighbour ferromagnetic and random Ising models, where frustration effects are not at play.

In this article, we perform continuous-time PQMC simulations of the two-dimensional quantum Edwards-Anderson (EA) Hamiltonian with Gaussian random couplings. A neural network state in the form of a self-learned restricted Boltzmann machine (RBM) is used as guiding wave function.
As we numerically demonstrate,  an RBM with a workable number of hidden neurons suppresses the systematic biases possibly affecting the energy and the replica spin overlap below sufficiently small statistical uncertainties, at least up to a hundred spins. The spin-overlap estimator is implemented here in PQMC simulations performing a single imaginary-time evolution of a two-fold replicated Hamiltonian; the pure estimator is then obtained via the standard forward walking technique.
A finite-size scaling analysis of the mean-squared EA order parameter and the corresponding Binder cumulant is performed. Corrections to the scaling ansatz turn out to be negligible for feasible spin numbers. This allows us to determine with fair accuracy the critical transverse field where the spin-glass quantum phase transition occurs, as well as the critical exponents of the correlation length and spin-glass susceptibility. For the latter two, good agreement is found with recent extreme-scale PIMC simulations~\cite{king2023quantum,bernaschi2023quantum} of models in the same universality class but featuring different random couplings. 
We also address the spin-overlap distribution within the spin glass phase. We observe that, at a transverse field approximately $20\%$ lower than the critical value, for the feasible sizes it displays a non-trivial structure with two (symmetry related) peaks and also substantial weight at zero overlap. The latter feature, when confirmed in the thermodynamic limit, is associated with replica symmetry breaking in classical systems~\cite{PhysRevLett.50.1946,PhysRevLett.64.1859,EnzoMarinari1999,marinari2000replica,PhysRevB.67.134410,PhysRevE.103.062111}.
To favour future computational studies on short-range quantum spin glasses, we provide via the repository of Ref.~\cite{brodoloni_2024_11534964} a  dataset of ground-state energies for 50 instances of the random couplings. These are useful to benchmark novel computational approaches for quantum spin models with frustrated interactions.\\

The rest of the Article is organized as follows:
Section~\ref{secmethod} defines the spin-glass Hamiltonian we study and it provides the essential details of the PQMC algorithm and of the RBM guiding wave function. The computation of the replica overlap is highlighted. The elimination of the population control bias  as a function of the random-walker population size and of the number of hidden neurons in the RBM is analyzed.
The quantum phase transition from the paramagnetic to the spin-glass phase is analyzed in Section~\ref{secresults}.
In Section~\ref{secconclusions} we summarize our main findings and we discuss some future perspectives.


\begin{figure}[h]
\centering
\includegraphics[width=1.\columnwidth]{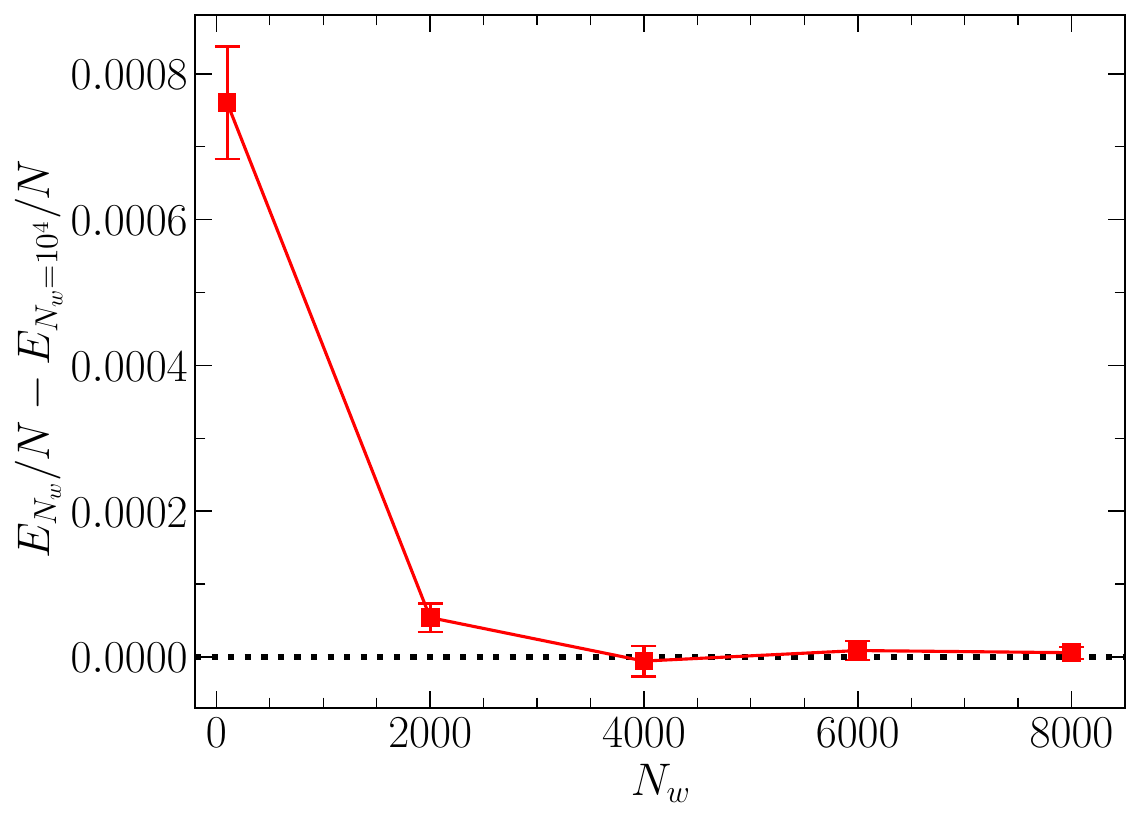}
\caption{
Bias in the energy per spin $E_{N_w}/N$ as a function of the average number of random walkers $N_w$, averaged over 5 instances of the 2D Gaussian EA model of size $L=10$, at transverse field $\Gamma=1.8$.
The reference energy is the results with $N_w=10^4$ walkers.
The RBM guiding wave-function features $N_h=128$ hidden neurons.
}
\label{fig1}
\end{figure}

\section{Models and Methods}
\label{secmethod}
\subsection{Two-dimensional Edwards-Anderson Hamiltonian with transverse field}
Our goal is to study the ground-state properties of the two-dimensional quantum EA model. For simulations performed with a single replica of the system, the Hamiltonian reads:
\begin{equation}
\hat{H}=-\sum_{\left<i,j\right>} J_{ij}{\sigma}^{z}_{i} {\sigma}^{z}_{j} -\Gamma \sum_{i=1}^{N} {\sigma}^{x}_{i}.
\label{H}
\end{equation}
$\sigma^x_{i}$ and $\sigma^z_{i}$ are conventional Pauli matrices at the lattice sites $i=1,\dots,N$, and $N=L^2$ is the  number of spins (in the single replica), with $L$ the (adimensional) linear system size. The triangular brackets $\left<i,j\right>$ indicate that the summation is performed over the nearest-neighbor nodes of a square lattice. The couplings $J_{ij}$ are randomly sampled from a Gaussian distribution with zero mean and unit variance, denoted as $\mathcal{N}(0,1)$.
$\Gamma$ is the intensity of the uniform transverse magnetic field. 
Hereafter, the eigenstates of the Pauli matrix ${\sigma^z_{i}}$ with eigenvalues $x_{i}=\pm1$ are denoted as $\left| x_{i} \right>$. 
%
%
A convenient computational basis is formed by states of $N$ spins $\left|\X \right> = \left| x_{1}  ... x_{N}\right>$, with $\X=(x_{1},\dots,x_{N})$. With $\left| \psi \right>$ we denote the state corresponding to the wave-function $\left< \X \right| \left. \psi \right> = \psi(\X)$.

\begin{figure}[h]
\centering
\includegraphics[width=1.\columnwidth]{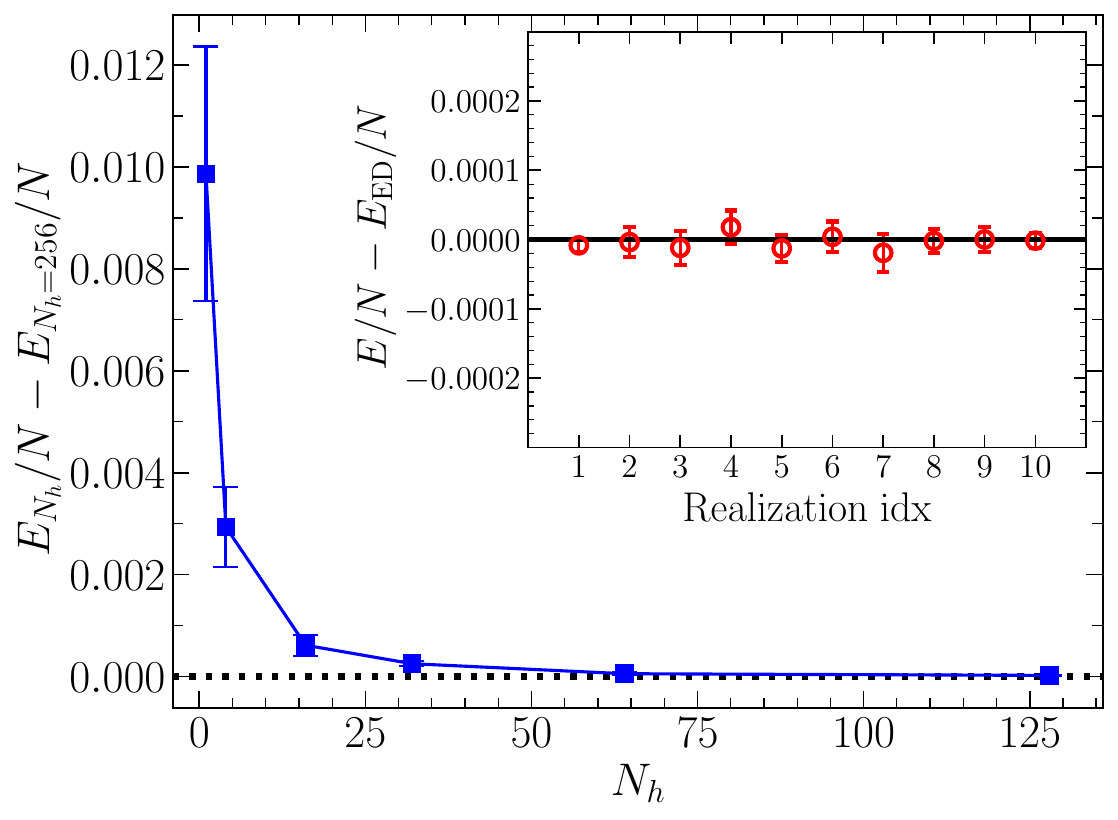}
\caption{
Main panel: Bias in energy per spin $E_{N_h}/N$ as a function of the number of hidden neurons $N_h$ in the RBM guiding-wave function, averaged over 5 instances as in Fig.~\ref{fig1}.
The reference energy is the results with $N_h=256$ neurons. 
The average number of random walkers is $N_w=4000$.
Inset: difference between PQMC predictions $E/N$ and exact-diagonalization results $E_{\mathrm{ED}}/N$ for 10 realizations of size $L=5$, labeled by the index $\mathrm{idx}=1,2,\dots,10$.
}
\label{fig2}
\end{figure}

\subsection{PQMC algorithm guided by neural network states}
To simulate the ground-state properties of the Hamiltonian~\eqref{H} we adopt a PQMC algorithm. A similar projection method was previously employed to simulate the annealing dynamics in Ref.~\cite{PhysRevE.75.036703}. Another zero-temperature method, but based on the stochastic series expansion algorithm, has recently been applied to the random-field quantum Ising model with ferromagnetic couplings~\cite{kramer2024quantumcritical}.
PQMC algorithms stochastically simulate the imaginary time  Schr\"odinger equation.
This allows projecting out the ground-state wave function $\psi_{0}(\X)$, starting from  an initial state $\psi(\X,0)$ at imaginary time $\tau =0$ (assumed not orthogonal to $\psi_{0}(\X)$), as follows: 
\begin{equation}
    \psi_{0}(\X) = \lim_{\tau \to \infty} \exp\left[-\tau\left(\hat{H}-{E_{\mathrm{ref}}}\right)\right] \psi(\X,0), 
\end{equation}
where $E_{\mathrm{ref}}$ is a reference energy introduced to stabilize the numerics, as explained below.
To improve efficiency, one introduces a guiding wave function $\psi_{g}(\X)$, namely, a suitable approximation for the ground state. The aim is to sample computational basis states proportionally to the (unnormalized) distribution $f({\X},\tau)=\psi ({\X},\tau) \psi_{g}({\X})$.
Long imaginary times are reached iterating many short time-steps $\Delta \tau$, according to the update rule:
\begin{equation}
\label{masterf}
f(\X,\tau+\Delta \tau) = \sum_{\X^\prime} \tilde{G}(\X,\X^\prime,\Delta \tau) f(\X^\prime,\tau), 
\end{equation}
where $\tilde{G}({\X},{\X}^\prime,\Delta \tau)= G(\X,\X^\prime,\Delta \tau)\frac{\psi_g({\X})}{\psi_g({\X^\prime})}$, and $G({\X},{\X}^\prime,\Delta \tau)= \left< \X \left| \exp\left[{-\Delta \tau(\hat{H}-E_{\mathrm{ref}})}\right] \right|\X^\prime \right>$ 
is the imaginary-time Green's function.
In this article, we adopt the continuous-time PQMC algorithm detailed in Refs.~\cite{PhysRevB.61.2599,becca2017quantum,PhysRevB.98.235145}, allowing us to avoid finite time-step errors.
Notice that continuous-time approaches are possible also for finite temperature simulations; see, e.g., the continuous Wolff algorithm of Ref.~\cite{PhysRevE.66.066110}.
The imaginary-time steps described by Eq.~\eqref{masterf} are implemented by evolving a population of random walkers, which undergo stochastic updates in the computational basis and a replication process, in jargon called branching, which accounts for the normalization of the update rule. 
The total number of random walkers has to be controlled to match a target population size $N_w$. 
This is achieved by dynamically tuning $E_{\mathrm{ref}}$.
After a sufficiently long imaginary time and for sufficiently large $N_w$, the walkers sample the elements of the computational basis proportionally to $f(\X,\tau) \underrel{\tau\to \infty}{=} \psi_{0}(\X)\psi_{\mathrm{g}}(\X)$. $f(\X,\tau)$ is assumed to be non-negative, consistently with the absence of a negative sign problem. As already noted, for finite $N_w$ the sampling might be biased.
Importantly, this bias can be totally suppressed with an appropriate choice of $\psi_g(\X)$. In fact, it can be shown that $N_w=1$ suffices if $\psi_g(\X)=\psi_0(\X)$.
Expectation values of  operators $\hat{O}$ are estimated via Monte Carlo integration. For general operators, one obtains the so-called mixed estimator $\left<\hat{O}\right>_{\mathrm{mixed}}=\frac{\left<\psi_{0}\right|\hat{O}\left|\psi_{g}\right>}{\left<\psi_{0}\left|\right.\psi_{g}\right>}$. If $\hat{O}=\hat{H}$ or if $\hat{O}$ and $\hat{H}$ commute, the mixed estimator coincides with the pure ground-state expectation value 
$\left<\hat{O}\right>=\frac{\left<\psi_{0}\right|\hat{O}\left|\psi_{0}\right>}{\left<\psi_{0}\left|\right.\psi_{0}\right>}$~\cite{RevModPhys.73.33}.
If this is not the case, the mixed estimator is biased when $\psi_{\mathrm{g}}(\X) \ne \psi_{\mathrm{0}}(\X)$. 
Clearly, this bias can be reduced  with an accurate choice of the guiding wave function. In the following, we adopt neural network states, which allow us to systematically improve $\psi_{\mathrm{g}}(\X)$  by increasing the number of hidden neurons in the RBM.
Furthermore, for diagonal operators in the computational basis, namely, if $\hat{O}=O(\X)$, the residual bias of the mixed estimator can be removed via the standard forward walking technique~\cite{PhysRevB.52.3654,boronat}. Again, an accurate guiding function shortens the required forward propagation time, allowing  obtaining for large $N_w$ the unbiased ground-state expectation value. 

Neural network states can be formed using energy-based generative neural networks  in the form of RBMs~\cite{doi:10.1126/science.aag2302}. 
These models define an unnormalized probability distribution over the spin variables $\X$, as follows: 
$P_{\mathrm{RBM}}(\X)  \propto \exp\left(\sum_i a_i x_i\right) \prod_m F_m(\X)$, where $F_m(\X)=2\cosh\left[b_m + \sum_{i=1}^N w_{im} x_i\right]$, where $m=1,\dots,N_h$ labels the $N_h$ neurons in the (unique) hidden layer of the RBM, while the weights $w_{im}$ and the biases $a_i$ and $b_m$ are determined by training the RBM.
By setting $a_i=b_m=0$ we enforce the $\mathbb{Z}_2$ symmetry.
Here, we adopt an unsupervised learning protocol, following Ref.~\cite{PhysRevE.100.043301}. The RBM is trained by minimizing the Kullback-Leibler divergence with respect to the walkers distribution. 
For this, the standard $k-$step contrastive divergence algorithm is employed~\cite{ACKLEY1985147}.
A sequence of PQMC simulations is performed. In each simulation, a large dataset of random walkers is accumulated, and this dataset is used to train an RBM. In every simulation, the guiding wave function $\psi_g(\X)=\sqrt{P_{\mathrm{RBM}}(\X)}$ based on the RBM trained after the previous run is used. Notice that this ansatz is legitimate since the ground-state wave function can be assumed to be real and nonnegative. 
Indeed, the Hamiltonian~\eqref{H} features non-positive off-diagonal elements in the chosen computational basis -- such models are sometimes referred to as stoquastic Hamiltonians~\cite{bravyi} --  allowing one to apply the Perron-Frobenius theorem.
The resulting self-learning protocol allows us to systematically improve the guiding wave function, thus reducing the population control bias that might occur for finite $N_w$.
For the typical simulation performed for this Article, each RBM is trained on datasets featuring $\simeq 1.2 \times10^{5}$ walker configurations. 
The one-step contrastive divergence algorithm is iterated for $3000$ steps using mini-batches of $40$ configurations.
We iterate PQMC runs followed by RBM training $\simeq 5$ times.
A long PQMC simulation is then performed with the optimal RBM. 
The computer times required for the whole process, considering, e.g., the results discussed in Fig.~\ref{fig1}, range from $\simeq 0.25$ to $\simeq 15$ hours using, e.g., 6 cores of an Intel(R) Xeon(R) Gold 6154 CPU processor, depending on $N_w$.

The roles of the walker-population size $N_w$ and of the hidden neuron number $N_h$ are analyzed in Fig.~\ref{fig1} and in Fig.~\ref{fig2}, respectively. The average energy per particle  $E/N=\left<\hat{H}\right>/N$ of five representative instances of the Hamiltonian~\eqref{H} is shown as a function of $N_w$ and $N_h$. The convergence to the unbiased asymptotic limits, which we identify with the cases $N_w=10^4$ (at $N_h=128$) and $N_h=256$ (at $N_w=4000$), is rapid. 
This demonstrates that the population control bias can be reduced well below the (modest) statistical uncertainties.
To further benchmark the neural PQMC algorithm, we make comparison against an alternative computational approach, namely, the exact diagonalization (ED) algorithm. This is practical up to system sizes $N\approx 30$.
In the inset of Fig.~\ref{fig2}, the discrepancy between the PQMC and the ED predictions for the ground-state energy per spin is shown, considering 10 Hamiltonian realizations of size $L=5$. We find vanishing values within the (small) statistical uncertainties, confirming the absence of bias in the PQMC predictions.

\begin{figure}[h]
\centering
\includegraphics[width=1.\columnwidth]{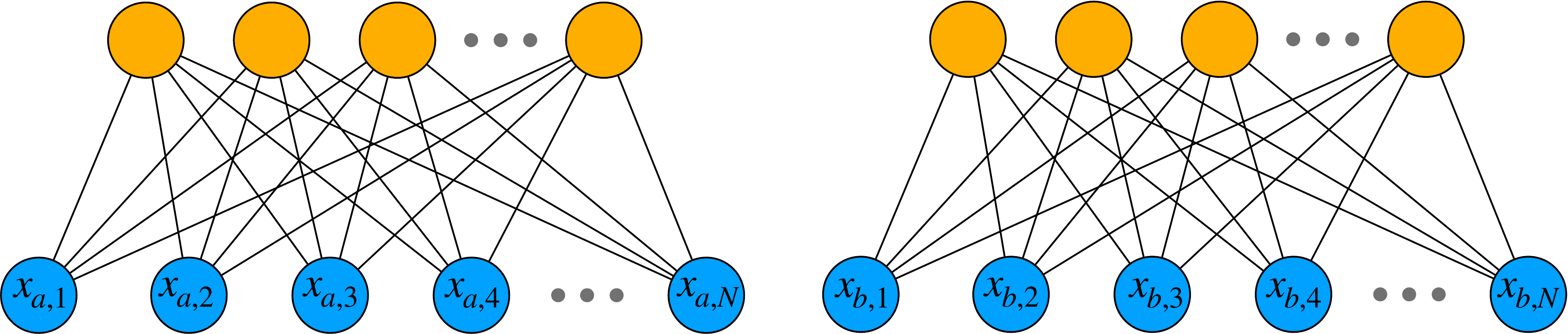}
\caption{
Representation of the connectivity of the RBM adopted to simulate the two-fold replicated EA Hamiltonian~\eqref{H2}.
The two sets of spins $x_{a,1},\dots,x_{a,N}$ and $x_{b,1},\dots,x_{b,N}$ (blue circles in the bottom) are connected to two corresponding sets of hidden variables (orange circles in the top).
The two replicas feature the same random couplings $J_{ij}$. 
}
\label{fig3}
\end{figure}

\subsection{Replica overlap in PQMC simulations}
To discern the spin-glass phase from the paramagnetic phase, we analyze the replica spin overlap. To determine this quantity in a PQMC simulation, we simulate a two-fold replicated system formed by the union of two identical copies of the EA model defined in Eq.~\eqref{H}. This replication leads to the following Hamiltonian of $2N$ spins:
\begin{equation}
\label{H2}
    \hat{H_2} = \sum_{\alpha=a,b} \left[ - \sum_{\left<i,j\right>} J_{ij} \sigma_{\alpha,i}^z \sigma_{\alpha,j}^z -\Gamma \sum_{i} \sigma_{\alpha,i}^x \right];
\end{equation}
here, $\sigma_{\alpha,i}^z$ and $\sigma_{\alpha,i}^x$ indicate standard Pauli matrices acting on spin $i$ of replica $\alpha=a,b$. 
It is worth emphasizing that the two replicas feature the same random couplings $J_{ij}$ and transverse field $\Gamma$. Furthermore, the whole Hamiltonian  is stochastically evolved in imaginary time using a unique random walker population.
Due to the absence of inter-replica interactions, the ground-state wave function of $\hat{H}_{2}$ must be separable: $\Psi_{\mathrm{0}}(\X_{a},\X_{b})= \psi_{\mathrm{0}}(\X_{a}) \psi_{\mathrm{0}}(\X_{b})$. A separable guiding wave function is conveniently implemented via an RBM featuring separable inter-layer connectivity, as shown in Fig.~\ref{fig3}.

\begin{figure}[h]
\centering
\includegraphics[width=1.\columnwidth]{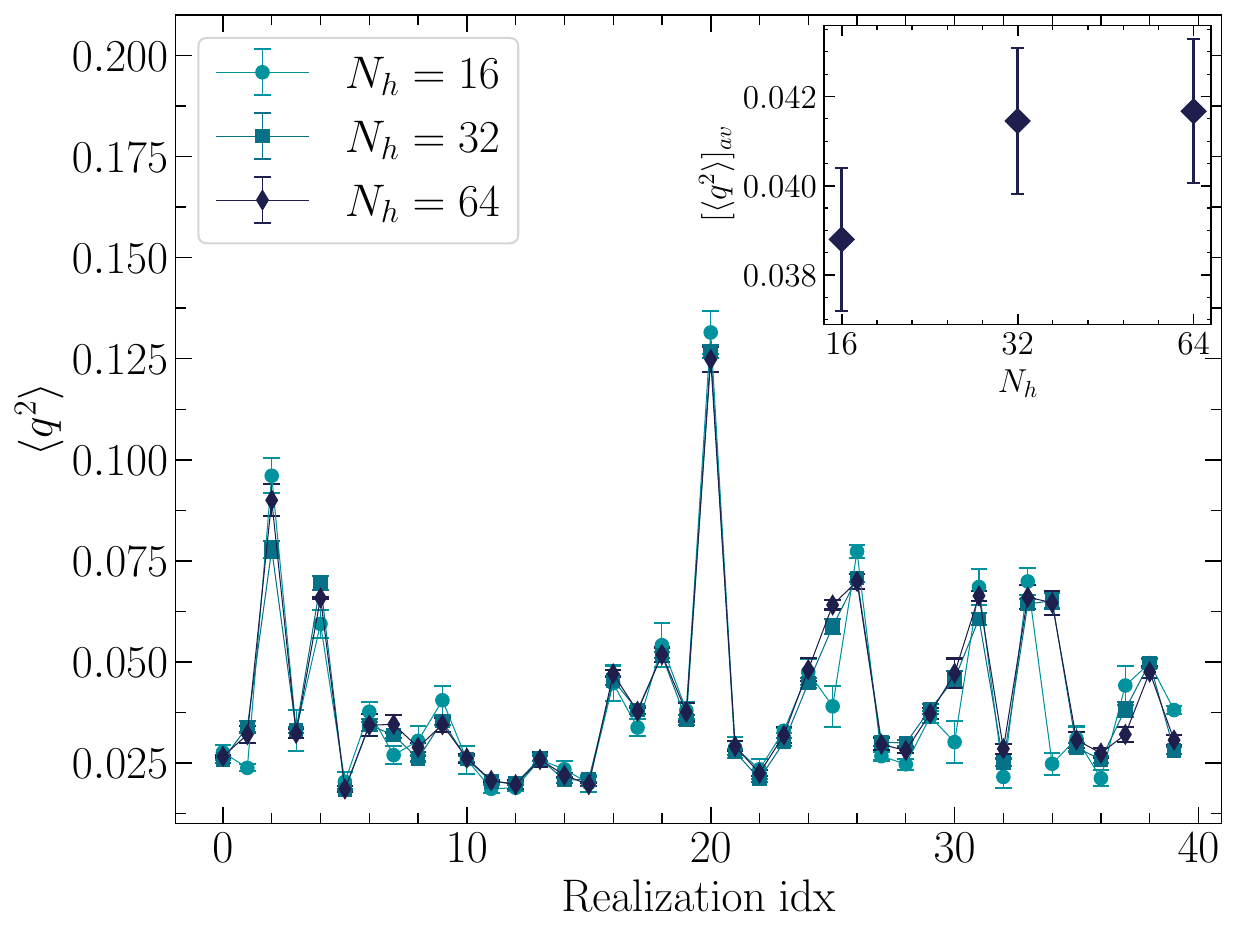}
\caption{
Mean-squared replica overlap $\left<q^2\right>$ for 40 instances of the 2D Gaussian EA model of size $L=10$, at transverse field $\Gamma=1.8$. The three datasets correspond to RBM guiding wave-functions with different numbers of hidden neurons $N_h$. The inset shows the average over 256 realizations as a function of $N_h$. 
}
\label{fig4}
\end{figure}

The spin-overlap operator is defined as: 
\begin{equation}
    \hat{q} = \frac{1}{N} \sum_{i=1}^N \sigma_{a,i}^z \sigma_{b,i}^z.
\end{equation}
This operator is diagonal in the chosen computational basis. Thus, we are able to estimate via forward walking the pure ground-state expectation value of the squared overlap $\left< q^2\right>$, namely, the mean-squared EA order parameter. This quantity allows discerning the spin glass from the paramagnetic phase. The disorder  average, obtained as the average over a large ensemble of $N_r$ realizations of the random couplings $J_{ij}$, will be indicated with square parenthesis: $\left[\left< q^2\right>\right]=  \chi_{\mathrm{sg}}/N$, where
$\chi_{\mathrm{sg}}$ indicates the spin-glass susceptibility.
It is important to inspect whether the RBM wave function allows us to suppress the population control bias. In Fig.~\ref{fig4}, we show $\left< q^2\right>$ for several representative instances of the Hamiltonian~\eqref{H} for $L=10$, obtained with different numbers $N_h$ of hidden neurons. The inset displays the average over an ensemble of 256 disorder realizations. One notices that the results for $N_h\ge 32$ are well within the statistical uncertainties, both for the ensemble average and, more stringently, for individual realizations. This indicates that possible systematic biases are under control.

In the study of the quantum critical point, it is convenient to consider the Binder ratio~\cite{PhysRevLett.72.4141,Alvarez}:
\begin{equation}
\label{eqR}
    R= \left[\frac{\left< q^4\right>}{\left< q^2\right>^2} \right].
\end{equation}
Notice that the Monte Carlo estimate of $R$ is possibly biased, since one evaluates a square and a ratio of expectation values approximated via Monte Carlo integration performed over a finite number of samples, say, $N_{\mathrm{MC}}$.  Still, as discussed in Ref.~\cite{king2023quantum}, this bias is expected to vanish as $N_{\mathrm{MC}}^{-1}$, i.e., faster than the statistical uncertainty, which scales as $N_{\mathrm{MC}}^{-1/2}$. A reduced bias~\cite{king2023quantum} is expected for the following alternative definition~\cite{doi:10.1142/97898128194370003}:
\begin{equation}
    R^{\prime} = \frac{\left[\left< q^4\right>\right]}{\left[\left< q^2\right>\right]^2}.
\end{equation}
As pointed out previously~\cite{PhysRevE.92.042107}, we also find that the latter definition leads to consistent results as with the definition of Eq.~\eqref{eqR}, but with enlarged fluctuations due to disorder averaging. See also the discussion in Ref.~\cite{kramer2024quantumcritical}.
Therefore, the results presented in the following Section have been computed with Eq.~\eqref{eqR}. Following the common convention, they are cast in the form of the Binder cumulant, defined as
\begin{equation}
\label{eqU}
    B=(3-R)/2.
\end{equation}

To inspect the nature of the spin-glass phase, it is instrumental to determine the spin-overlap distribution, defined, for $q\in[-1,1]$ and for a given choice of the random couplings, as:
\begin{equation}
    P_J(q) = \left< \delta(q-\hat{q}) \right>.
\end{equation}
In practice, we determine a discrete histogram with $N$ bins.
The ensemble-average spin-overlap distribution $P(q)=\left[P_J(q) \right]$ is the mean over $N_r$ realizations of the couplings.

\section{Results: quantum spin glass transition}
\label{secresults}
The 2D EA Hamiltonian with transverse field is expected to host a quantum phase transition from a paramagnetic to a spin-glass phase when the transverse field is reduced below a critical value. It is worth mentioning here that, in the classical EA model, the spin-glass phase only occurs at zero temperature~\cite{PhysRevE.84.046706}. This is in contrast with the corresponding 3D model or the SK Hamiltonian, which host a classical spin-glass transition at a finite critical temperature.
For the quantum SK model, the occurrence of replica symmetry breaking at sufficiently small transverse field was recently proven~\cite{PhysRevLett.127.207204}.
Hereafter, we analyze the quantum phase transition in the 2D quantum EA Hamiltonian with Gaussian random couplings. Very recent extreme-scale PIMC simulations investigated the analogous quantum phase transition considering binary random couplings~\cite{king2023quantum,bernaschi2023quantum}.

\begin{figure}[h]
\centering
\includegraphics[width=1.\columnwidth]{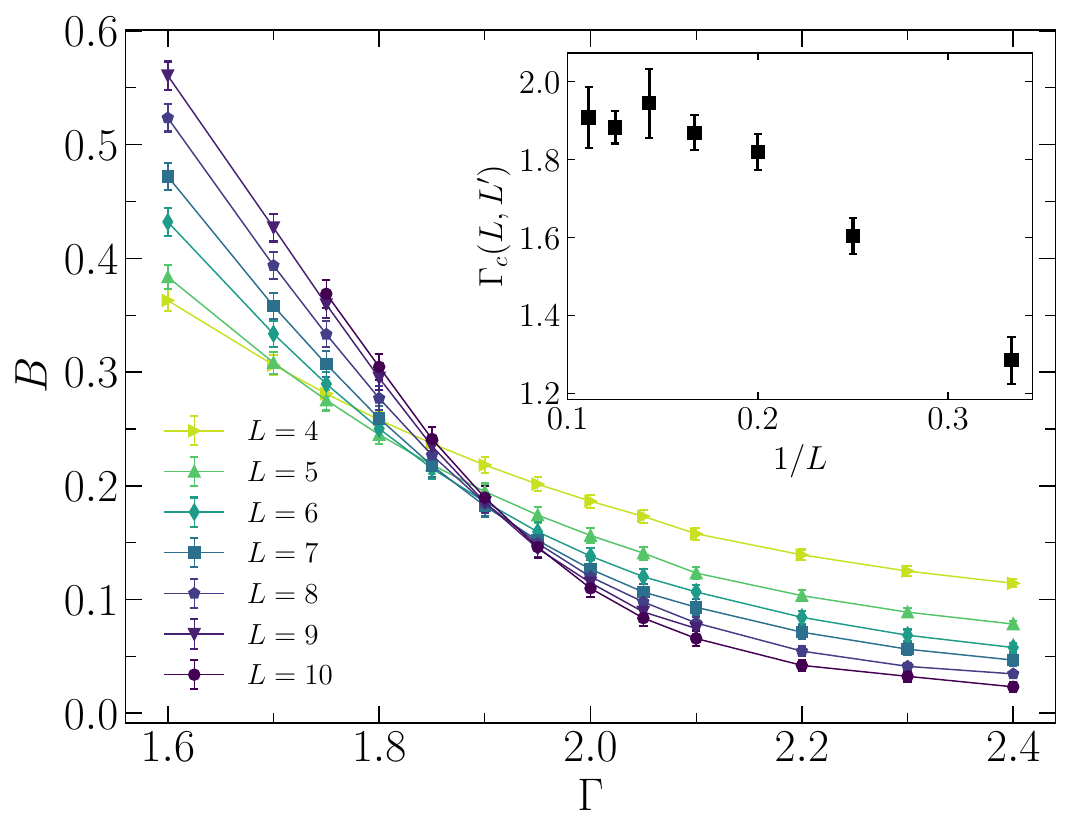}
\caption{
Binder cumulant $B$ as a function of the transverse field $\Gamma$. Different curves correspond to different system sizes $L$.
The inset shows the crossing point $\Gamma_c(L,L^\prime)$ of consecutive system sizes $L$ and $L^\prime$, as a function of the (smaller) system size $L$.
}
\label{fig5}
\end{figure}

First, we analyze the finite-size scaling of the (realization averaged) Binder cumulant $B$ of the mean-squared EA order parameter, defined in Eq.~\eqref{eqU}.
This quantity is expected to be independent of the system size at the critical point, apart from corrections to the universal scaling ansatz. The latter is expected to hold when approaching the thermodynamic limit, but finite-size corrections might be sizable for small $L$. This behaviour is indeed what we observe; see results in Fig.~\ref{fig5}. These data are averaged over a number of realizations of the random couplings ranging from $N_r=256$ to $N_r=1536$.
Due to the finite-size effects, the crossing points of consecutive system sizes drift to larger transverse fields. However, for $L\ge 6$, the crossing points saturate within statistical uncertainties. 
In the large $L$ regime, the crossings occur around the transverse field $\Gamma_c \approx 1.9$. This represents a first approximate estimate of the quantum critical point.

\begin{figure}[h]
\centering
\includegraphics[width=1.\columnwidth]{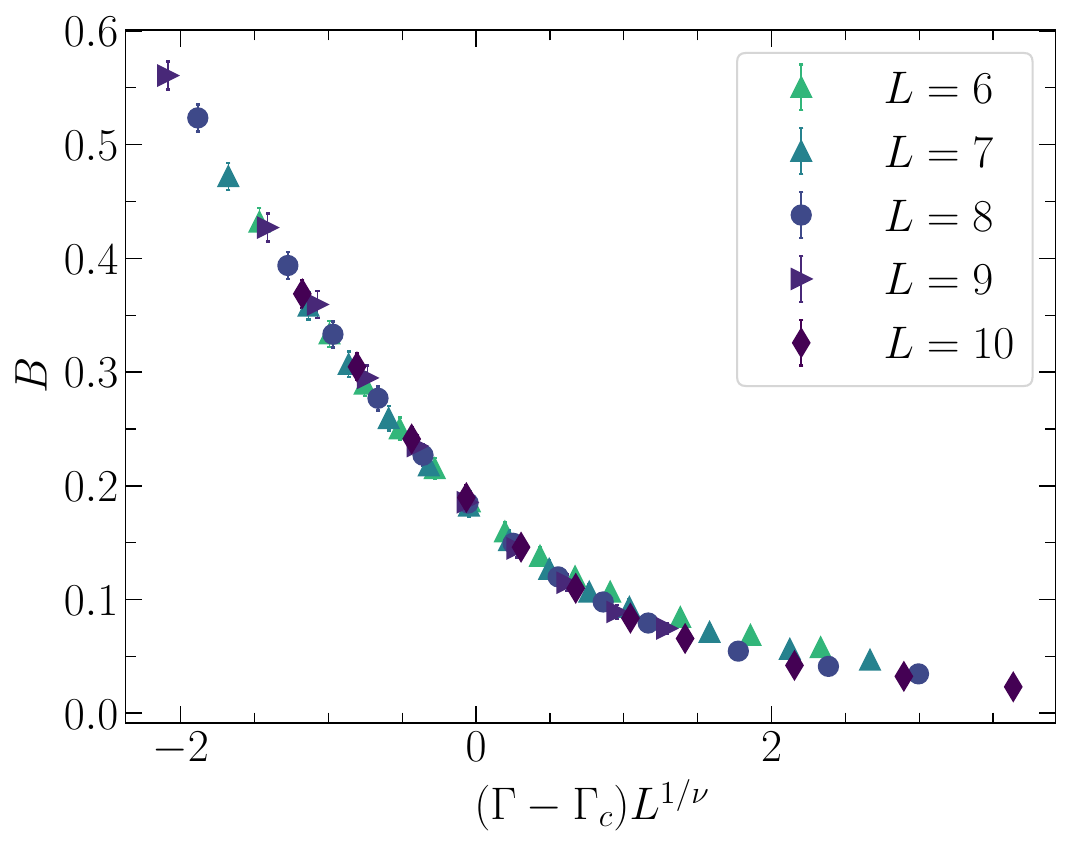}
\caption{
Collapse of the Binder cumulant $B$ as a function of the rescaled transverse field $(\Gamma-\Gamma_c)L^{1/\nu}$. $\nu$ denotes the critical exponent of the correlation length. Different symbols correspond to different system sizes $L$.
}
\label{fig6}
\end{figure}

The Binder cumulant is expected to scale as $B=f_B\left( (\Gamma-\Gamma_c) L^{1/\nu} \right)$, where $\nu$ is the critical exponent of the correlation length, $\Gamma_c$ is the critical transverse field, and $f_B(x)$ is a universal scaling curve. $\Gamma_c$ and $\nu$ are obtained from a best-fit analysis. In this analysis, we expand the scaling function as $f_B(x) = f_{B0}+f_{B1}x+f_{B2}x^2$, treating also $f_{B0}$, $f_{B1}$, and $f_{B2}$ as fitting parameters. The collapse of the data corresponding to different system sizes is indeed verified in Fig.~\ref{fig6}, considering the sizes $L\ge 6$.

\begin{figure}[h]
\centering
\includegraphics[width=1.\columnwidth]{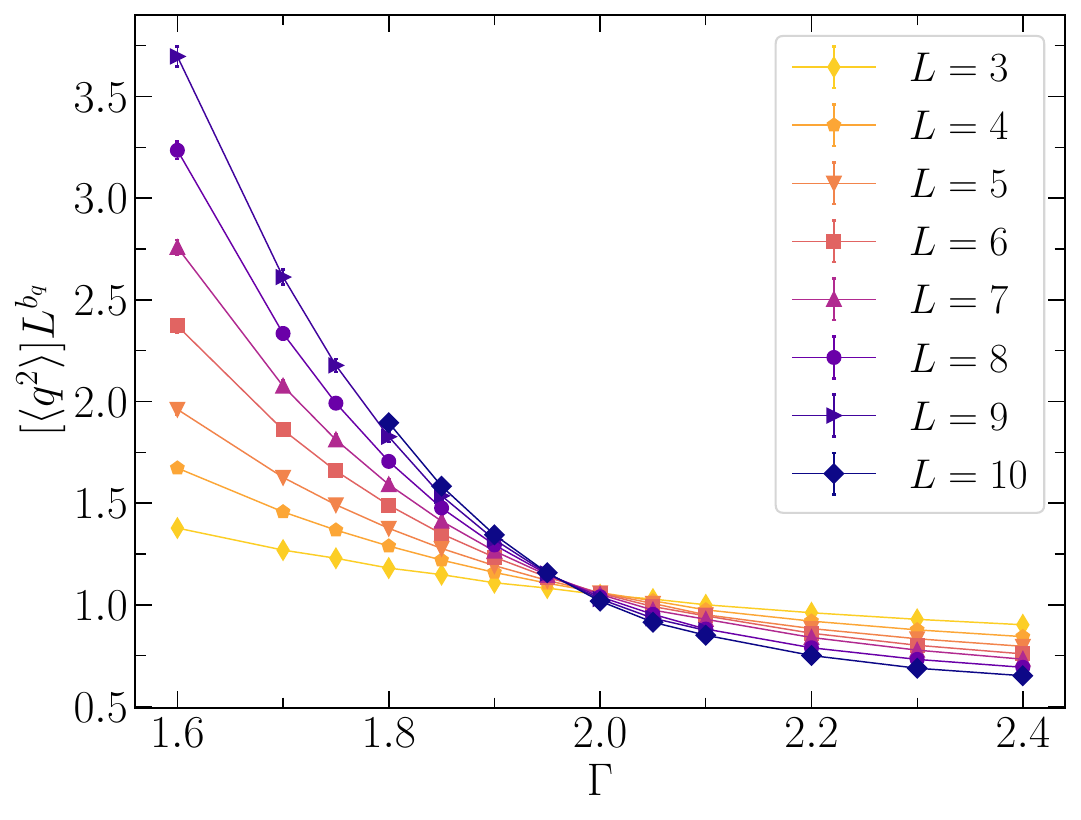}
\caption{
Rescaled (disordered averaged) EA order parameter $\left[\left<q^2\right>\right]L^{b_q}$ as a function of the transverse field $\Gamma$. 
$b_q$ is the critical exponent associated with $\left<q^2\right>$.
Different datasets correspond to different system sizes.
}
\label{fig7}
\end{figure}

\begin{figure}[h]
\centering
\includegraphics[width=1.\columnwidth]{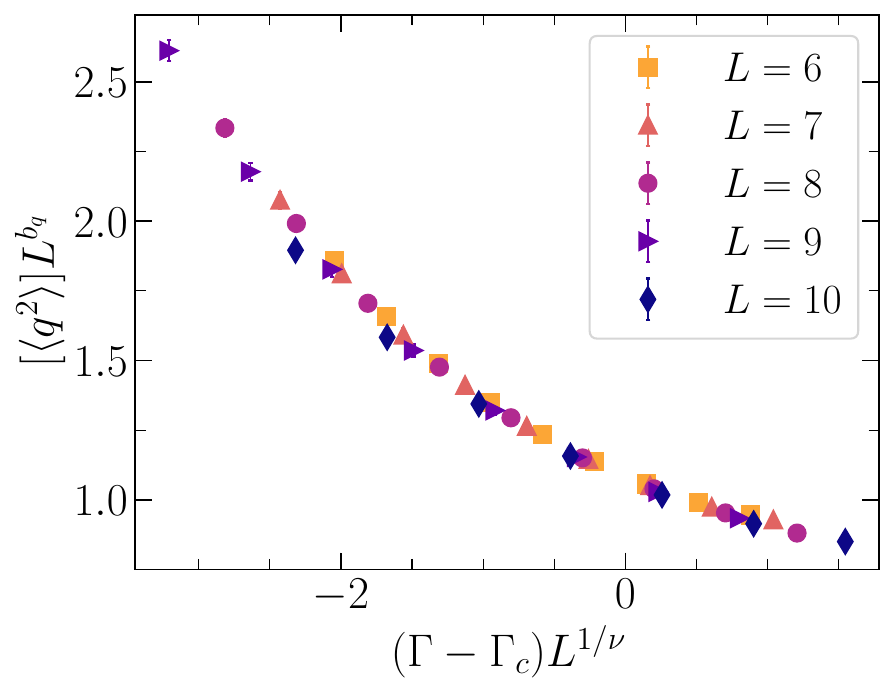}
\caption{
Collapse of the rescaled (disordered averaged) EA order parameter $\left[\left<q^2\right>\right]L^{b_q}$ as a function of the rescaled transverse field $(\Gamma-\Gamma_c) L^{1/\nu}$. 
}
\label{fig8}
\end{figure}

To pinpoint the critical transverse field more precisely, we consider the (realization averaged) mean-squared EA order parameter $\left[\left<q^2\right>\right]$. Notice that this coincides with $\chi_{\mathrm{sg}}/N$. Its finite-size scaling ansatz is $\left[\left<q^2\right>\right]L^{b_q}=f_q\left( (\Gamma-\Gamma_c) L^{1/\nu} \right)$, where $b_q$ is the corresponding critical exponent. Accordingly, the rescaled quantity $\left[\left<q^2\right>\right]L^{b_q}$ should be system size independent at the critical point $\Gamma_c$. As shown in Fig.~\ref{fig7}, data corresponding to different system sizes cross essentially at the same value of the transverse field, even for system sizes as small as $L\simeq 3$. This suggests that the corrections to the scaling ansatz are less pronounced here than in the case of the Binder cumulant $B$, allowing us to better estimate $\Gamma_c$. Still, to avoid finite size biases, in the data collapse we consider sizes $L\ge 6$. The scaling function is expanded as $f_q(x)=f_{q0}+f_{q1}x+f_{q2}x^2$, and  the best-fit analysis is performed using the software available from Ref.~\cite{melchert2009autoscalepy}.
The data collapse is verified in Fig.~\ref{fig8}.
Importantly, the best-fit analysis provides estimates of $\nu$, of $b_q$, and of the critical point $\Gamma_c$. For the latter, we find $\Gamma=1.98(7)$, consistently with the findings based on $B$ obtained in the large $L$ regime (see inset of Fig.~\ref{fig5}). The two critical exponents are expected to be universal. In fact, we find good agreement with recent estimates, obtained via PIMC simulations in the zero-temperature limit, based on different choices of the random couplings. 
For the correlation-length critical exponent, we find $1/\nu=1.11(22)$. Within the reported error bar, this results is consistent with the bound $\nu\ge 2/D$, where $D$ is the dimensionality~\cite{PhysRevLett.57.2999}.
Our results, as well as the corresponding data from the recent literature, are summarized in Table~\ref{table1}. 
To determine the statistical uncertainties, we repeat the fitting process 55 times starting from randomized initial guesses. This allows computing the means of the standard deviations. Systematic errors due to scaling corrections are estimated considering the fluctuations obtained excluding the $L=6$ size and then also the case $L=7$. The error bars we report in parenthesis correspond to the larger between the statistical and the systematic uncertainty.

\begin{table}[H]
\centering
\begin{tabular}{|c|c|c|c|c|}
\hline
Couplings           & $1/\nu$     & $b_q$   & $\Gamma_c$ & Ref.      \\ \hline
$50\%$. $\pm 1$     & 1.02(16)    & 1.76(3) & 2.11(1)    & \cite{king2023quantum}     \\
$50\%$ $\pm 1$     & 0.71(24)(9) & 1.73(8)(8)       & 2.18(1)    & \cite{bernaschi2023quantum}\footnote{Ref.~\cite{bernaschi2023quantum} reports 
 the critical value $k_c = 0.2905(5)$, with $\Gamma=-\frac{1}{2k}\log\tanh(k)$, and $\chi_{\mathrm{sg}}\propto L^{0.27(8)(8)}$. Also, for sizes close to ours, namely $L=8$ and $L=12$, that study reports  an effective exponent $1/\nu=1.07(7)$, indeed very close to our result.}    \\
$\mathcal{N}(0, 1)$ & 1.11(22)    & 1.68(8) & 1.98(7)    & This work \\ \hline
\end{tabular}
\caption{
Summary of the critical exponents of the correlation length $\nu$ and of the squared EA order parameter $b_q$, as well as of the critical transverse fields of the spin-glass transitions $\Gamma_c$, corresponding to different choices of the random couplings $J_{ij}$ (reported in the first column). The last column reports the reference. When two errors are reported in parentheses, the first corresponds to the systematic error, the last to statistical uncertainty.
}
\label{table1}
\end{table}

\begin{figure}[h]
\centering
\includegraphics[width=1.\columnwidth]{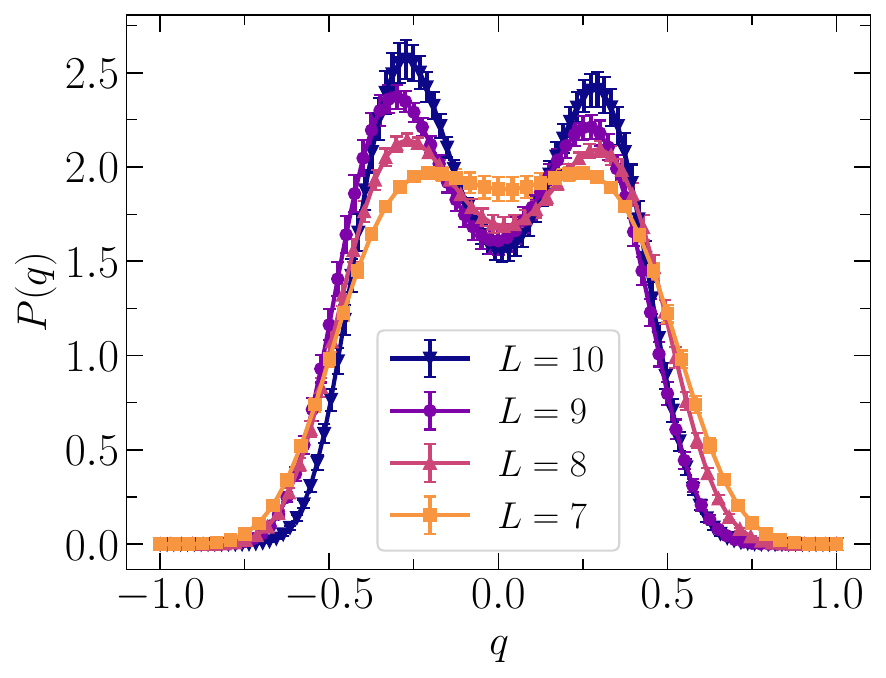}
\caption{
Disorder averaged replica spin overlap distribution $P(q)$. Different datasets correspond to different sizes $L$. The transverse field is $\Gamma=1.6<\Gamma_c$.
}
\label{fig9}
\end{figure}

To inspect the nature of the spin-glass phase, we perform an exploratory analysis of the zero-temperature replica spin-overlap distribution $P(q)$ for $\Gamma < \Gamma_c$. Specifically, we set $\Gamma = 1.6$, and we consider the system sizes $L=7,8,9,\text{ and }10$. The histograms, which are averaged over a number of realizations ranging from $N_r=1186$ to $N_r=1931$, are shown in Fig.~\ref{fig9}.
For relatively large $L$, the distributions display two sizable peaks, symmetrically located around $q \simeq \pm 0.25$. For the smallest size $L=7$ the peaks are significantly less pronounced. 
At this small $\Gamma$, the ground state is well within the spin-glass phase. This implies that the Monte Carlo autocorrelation times increase compared to the regime $\Gamma \simeq \Gamma_c$ considered above.
It is verified that  for $L=7$  essentially all disorder realizations satisfy the $\mathbb{Z}_2$ symmetry, corresponding to the invariance $P_J(q)=P_J(-q)$. 
For the largest sizes, in particular for $L=10$, many simulations  randomly break the symmetry for the simulations times considered in this analysis. It is verified that, performing much longer simulations for representative disorder realizations, the symmetry is recovered. 
%
%
Also, when ensemble averaging is performed, the symmetry is fulfilled within statistical uncertainties. The latter are estimated from the fluctuations over a five-fold random splitting of the realization ensemble.
%
%
%
In the case of classical spin-glass models, special attention is devoted to the distribution weight at $q\simeq 0$. Two prominent theories of the spin-glass phase lead to different predictions. The droplet theory predicts vanishing values, namely $P(0)=0$, in the thermodynamic limit, while in the framework of replica symmetry breaking a finite value is expected (see, e.g., Refs.~\cite{marinari2000replica,PhysRevE.103.062111}). 
In the quantum model addressed here, large values are found for the workable system sizes, with a slow decrease as a function of $L$. 
Finite values $P(0)$ were also found in finite temperature PIMC simulations of the quantum SK model at weak transverse field~\cite{PikYinLai1990}.
However, it is worth pointing out that larger system sizes are needed to ascertain the correct asymptotic behaviour valid for $L \to \infty$. In fact, Refs.~\cite{PhysRevLett.81.4252,PhysRevB.62.946} argue that this is the case, in particular, when the overlap distribution is evaluated very close to the phase transition. Notice that this argument has been debated~\cite{PhysRevB.63.184422}. Performing PQMC simulations even deeper in the spin-glass phase is in principle possible, but it comes at the cost of further increased correlation times, which require longer simulation times for ergodic sampling and, thus, problematic computational costs.

\section{Conclusions}
\label{secconclusions}
We have investigated the spin-glass quantum phase transition in the 2D EA Hamiltonian with transverse field. Our study focused on the case of Gaussian couplings, complementing recent  investigations that addressed the case of binary random couplings~\cite{king2023quantum,bernaschi2023quantum}. Notably, we have shown that guiding wave functions in the form of self-learned neural network states allow removing the population control bias in PQMC simulations of frustrated quantum spin models. To compute the replica spin overlap, which gives us access to the mean-squared EA order parameter $\left[\left<q^2\right>\right]$, a two-fold replicated Hamiltonian has been evolved in imaginary time.
A finite-size scaling of $\left[\left<q^2\right>\right]$ allowed us to pinpoint the critical transverse field where the spin-glass quantum phase transition occurs, finding $\Gamma_c = 1.98(7)$. For the critical exponents, good agreement is found with the recent PIMC results for binary random couplings (see Table~\ref{table1}).
We also performed an exploratory analysis of the overlap distribution $P(q)$ in zero-temperature quantum spin glasses. For the feasible system sizes $N\lesssim 100$ and for a transverse field approximately $20\%$ lower than the critical value $\Gamma_c$, the function $P(q)$ displays a non-trivial shape. It features two (symmetry related) peaks, and substantial weight at $q\simeq 0$. The latter feature, when confirmed in the thermodynamic limit, is associated with replica symmetry breaking.
Notably, to favour the development of novel computational methods for quantum Ising models, we provided at the repository of Ref.~\cite{brodoloni_2024_11534964} a dataset of ground-state energies for an ensemble of disorder realizations.
Specifically, the dataset includes couplings and energies of 50 Hamiltonian instances at $\Gamma=1.8$ and system size $L=10$.
This dataset represent a useful benchmark also for quantum simulation platforms, e.g., for Rydberg-atom arrays.

Our study highlights the instrumental role of neural network states in simulating otherwise computationally overwhelming quantum many-body systems. Specifically, we have used them to implement an unbiased quantum Monte Carlo algorithm for the ground state of frustrated disordered quantum Ising models. This technique is complementary to, e.g., highly optimized  PIMC algorithms~\cite{BERNASCHI2024109101} or stochastic series expansion methods~\cite{kramer2024quantumcritical}.
Neural-network driven quantum Monte Carlo algorithms might also be useful to simulate the tunneling dynamics of quantum annealers~\cite{doi:10.1126/science.1068774,PhysRevA.93.032304,PhysRevLett.117.180402,PhysRevB.96.134305,PhysRevA.97.032307,PhysRevB.100.214303,andriyash2017can}. Such simulations could help understanding how adiabatic quantum computers solve hard optimization problems.
Future endeavours might focus on more experimentally relevant setups, e.g., on models for trapped-ion~\cite{smith2016many} or Rydberg-atom quantum computers~\cite{labuhn2016tunable,ebadi2021quantum}. In the latter platform, frustrated lattices, such as the kagome geometries, have been implemented, and the possibility to observe glassy phases has been discussed~\cite{PhysRevLett.116.135303,PhysRevLett.130.206501,PhysRevLett.132.206503,he2024floquet,hibat2024recurrent}.
In particular, positional disorder can be controlled essentially at will, allowing the implementation of, e.g., models for amorphous materials~\cite{juliafarre2024amorphous}. This paves the way to the investigation of the different localization properties of purely random noise, correlated disorder, and quasi-periodic patterns~\cite{PhysRevA.72.053607,PhysRevB.87.134202,PhysRevA.95.013613,PhysRevLett.124.130604,PhysRevB.99.165131,PhysRevLett.131.173402}.
Finally, it is worth mentioning that the neural PQMC algorithm could be further improved adopting tailored neural-network architectures or exploiting different training strategies~\cite{PhysRevE.101.063308,PhysRevE.101.053301,PhysRevB.107.165149, WU2024109169}.
The extension to Hamiltonians affected by the negative sign problem might also be possible using signed random walkers combined with walker annihilation techniques~\cite{10.1063/1.3193710}.

\section*{Acknowledgments} 
We acknowledge useful discussions with Alexandre Dauphin, Joseph Vovrosh, Sergi Julia Farr\'e, Simone Cantori, Giuseppe Scriva, Estelle Maeva Inack, Pierbiagio Pieri, Víctor Mart\'in-Mayor and Isidoro Gonz\'alez-Adalid Pemart\'in.
This work was supported by the PNRR MUR Project No. PE0000023-NQSTI and by the Italian MUR under the PRIN2022 project ``Hybrid algorithms for quantum simulators'' No. 2022H77XB7.
Partial contributions from the PRIN-PNRR 2022 MUR project ``UEFA'' - P2022NMBAJ are also acknowledged.
We acknowledge support from the CINECA ISCRA award IsCb2 NEMCASRA and the CINECA-INFN agreement, for the availability of high-performance computing resources and support.
We also acknowledge the EuroHPC Joint Undertaking for awarding this project access to the EuroHPC supercomputer LUMI, hosted by CSC (Finland) and the LUMI consortium through a EuroHPC Regular Access call.

\bibliography{bibliography}{}

\end{document}